# On the pressure dependence of the thermodynamical scaling exponent $\gamma$

R. Casalini[1,*] and T. C. Ransom[1,2]

[1]Naval Research Laboratory, Chemistry Division, Washington DC 20375-5342, USA.

[2]American Society for Engineering Education, Washington, D.C. 20036-2479, USA.

October 25 2019

**Abstract**

Since its initial discovery more than fifteen years ago, the thermodynamical scaling of the dynamics of supercooled liquids has been used to provide many new important insights in the physics of liquids, particularly on the link between dynamics and intermolecular potential. A question that has long been discussed is whether the scaling exponent $\gamma$ is a constant or it depends on pressure. Here we offer a simple method to determine the pressure dependence of $\gamma$ using only the pressure dependence of the glass transition and the equation of state.

Using this new method we find that for the six non-associated liquids investigated $\gamma$ always decreases with increasing pressure. Importantly in all cases the value of $\gamma$ remains always larger than 4. Liquids having $\gamma$ closer to 4 at low pressure show a smaller change in $\gamma$ with pressure. We argue that this result has very important consequences for the experimental determination of the functional form of the repulsive part of the potential in liquids.

[*] Corresponding author : riccardo.casalini@nrl.navy.mil

**Introduction**

The density and temperature dependence of dynamic properties of liquids and polymers (i.e. viscosity, relaxation and diffusion time) has been found to be well described by the thermodynamical scaling (TDS) behavior [1,2,3,4,5]

$$\log(X) = \Im(T\rho^{-\gamma}) \quad (1)$$

Where $X$ is a dynamic property (relaxation time, viscosity, etc), $\Im$ is an unknown function, $T$ the temperature, $\rho$ the density, and $\gamma$ the thermodynamical scaling exponent.

In the literature it has lengthily been debated whether the exponent of the thermodynamical scaling, γ, for nonassociated liquids is constant or state-point dependent.[6,7,8,9,10,11,12,13] It has been showed for many materials that a constant γ gives a very good superposition of various dynamic properties over a broad range of density and temperature. However, in a recent investigation we have found unequivocal evidence that dielectric relaxation data for a nonassociated liquid (DC704) cannot be scaled according to eq.(1) because the exponent is state-point dependent, decreasing with increasing pressure.[12] These deviations are in agreement with the predictions of the isomorph theory.[14, 15] Thus, it is of interest to re-analyze past results on other nonassociated liquids to investigate if a similar dependence can be found.

A standard way to test whether or not the parameter γ is constant is to plot $\ln(T)$ versus $\ln(\rho)$ at constant $X$. From eq.(1), if γ is a constant then the data should follow a linear behavior. Thus, the problem is reduced into determining the degree of deviation from a linear fit of a few points (typically 4) is significant. This is not trivial since the range of T and ρ are limited. Another problem of this method is also that it is not clear what should be the function describing the state point dependence of γ.

To overcome this problem in here, we re-analyze existent data using a different approach. Recently it was proposed to determine the state-point dependence of γ using the equation [9,12,16]

$$\gamma = \frac{\Delta V}{\kappa_T E_P - T \Delta V \alpha_P} \quad (2)$$

where $\Delta V \left(= RT \left(\frac{\partial \ln(X)}{\partial P}\right)_T\right)$ is the activation volume, $\kappa_T$ the isothermal compressibility, $E_P$ the isobaric activation energy and $\alpha_P$ is the isobaric expansion coefficient at atmospheric pressure. Using this method, we reported the determined the variation of γ with pressure for the liquid DC704, decreasing from $\gamma \approx 7$ at atmospheric pressure to $\gamma \approx 4$ at P= 0.9 GPa.[12]

Recently,[18] we have also shown that taking into consideration the available data for nonassociated liquids, out of fifty liquids only for two reported values of γ are smaller than 4 and both liquids are extremely polar, propylene carbonate ($\gamma = 3.7$, dipole moment μ $\cong$ 3.9D) and acetonitrile ($\gamma = 3.5$, μ $\cong$ 4.9D). Since molecular dynamic simulations have shown that a large dipole moment is expected to cause a decrease of $\gamma$, the polarity of these two liquids may explain their lower value of $\gamma$.[17] Thus, the value $\gamma \approx 4$ appear to be a limit behavior for nonassociated liquids.

Recently, we also showed [18] that eq.(2) can be further simplified to

$$\gamma = \frac{1}{T\left(\kappa_T \left.\frac{\partial P}{\partial T}\right|_X - \alpha_P\right)} \quad (3)$$

Using this equation, the state-point dependence of γ can be determined using just three quantities. Using Eq. (3), we also investigated three associated liquids: Glycerol, Dibutyl phthalate, and Dipropylene glycol, and found that the exponent γ increases (from γ<4) towards $\gamma \approx 4$ at high pressure.[18]

Here we present a simple derivation obtaining an analytical function for the state dependence of $\gamma$ instead of determining $\gamma$ at discrete experimental points using eq. (3) or by deviation from the linear behavior of $\ln(T_X)$ versus $\ln(\rho_X)$. We show how the pressure behavior of $\gamma$ can be deducted from the pressure behavior of the temperature at constant $X$, $T_X$.

Using this new method we present new data on the pressure behavior of the thermodynamical scaling exponent $\gamma$.

**Methods**

*Derivation of γ(P) equation.* The pressure dependence of the temperature at a fixed values of $X$, $T_X(P)$, has been found for several system to be non-linear, and its behavior can be described by the empirical equation of Andersson and Andersson (AA) [19]

$$T_X(P) = T_0 \left(1 + \frac{P}{P_0}\right)^{\frac{1}{a}} \qquad (4)$$

where $T_0$, $a$ and $P_0$ are constants. With the derivative $\frac{\partial T_X}{\partial P} = \frac{T_0}{aP_0}\left(\frac{P}{P_0}+1\right)^{\frac{1}{a}-1}$. This equation has been verified for a large number of materials by many different experimental groups.[20, 21, 22, 23, 24, 25, 26] Although the AA equation was originally introduced empirically, It has been also derived from theoretical models.[27,28]

The dependence of the density from pressure and temperature is well described by the Tait equation of state (EoS) [29].

$$\rho(T,P) = \rho_0(T)\left\{1 - C\ln\left[1 + \frac{P}{b_0 \exp(-b_1 T)}\right]\right\} \quad (5)$$

Where $\rho_0(T)$ is the temperature dependent density at zero pressure (described either as a polynomial or exponential) and $C$, $b_0$ and $b_1$ are constants.

By combining eq.(4) and eq.(5) we can describe the pressure dependence of the density at constant $X$, $\rho_X(P)$ as

$$\rho_X(P) = \rho_0(T_X(P))\left\{1 - C\ln\left[1 + \frac{P_0\left[\left(\frac{T_X(P)}{T_0}\right)^a - 1\right]}{b_0\exp(-b_1 T_X(P))}\right]\right\} \quad (6)$$

Rewriting the TDS condition as

$$\gamma = \frac{\partial \ln(\rho_X)}{\partial \ln(T_X)} \quad (7)$$

we can determine the pressure dependence of the exponent $\gamma$ as

$$\gamma(P) = \frac{\rho_X}{T_X}\frac{\partial T_X}{\partial \rho_X} \quad (8)$$

Therefore, from the behavior of the pressure dependence of $T_X$ (eq.(4)) together with an EoS, it is possible to determine the pressure dependence of the exponent $\gamma$ without the need to directly analyze the deviation from the linear behavior of $\log(\rho_X)$ versus $\log(T_X)$.

Substituting eq.(6) in eq.(8) it is possible to determine the analytical function of $\gamma(P)$

$$\gamma(P) = \left\{T_X(P)\left[-\alpha_P + \frac{C}{1 + \ln\left(1 + \frac{P}{B(P)}\right)}\frac{\left(\frac{a(P+P_0)}{T_X(P)} + b_1 P\right)}{P + B(P)}\right]\right\}^{-1} \quad (9)$$

Where

$$B(P) = b_0 \exp[-b_1 T_X(P)] \tag{10}$$

It is interesting to note that considering the typical values of the parameters for nonassociated liquids in eq.(9), the term due to $\alpha_P$ in eq.(9) (and eq.(3)) varies much less than the second term related to the compressibility; the latter increases with pressure, causing the decrease of γ. It is important to notice that an extrapolation to much higher pressure than the EoS data or the $T_X(P)$ data is likely to give unreasonable results, since the high pressure validity of the two starting equations is unknown.

Below we use this method for six nonassociated liquids for which the high pressure behavior of the dielectric relaxation time has been previously investigated. For these liquids a constant γ was found to give a good superposition of dynamic data, and the plots of $\log(\rho_X)$ versus $\log(T_X)$ are nearly linear, which would imply a nearly constant γ.

**Analysis of experimental results**

The dielectric relaxation and EoS data as well as γ exponent were previously published for six non-associated liquids: o-terphenyl (OTP), γ=5.3,[30,31] 1,1'-di(4-methoxy-5-methylphenyl)cyclohexane (BMMPC), γ=8.5,[32,33], Phenylphthalein-dimethylether (PDE), γ=4.5,[34,35] and three polychlorinated byphenyls (PCB42, PCB54 and PCB62), found to have very different values of γ (PCB42 γ=5.5, PCB54 γ=6.7 and PCB62 γ=8.5).[36] In particular, between these materials BMMPC and PCB62 have been reported to have some of the largest values of γ reported in the literature for dielectric relaxation data. Dielectric relaxation spectroscopy data were used in this study because they have the advantage of a larger frequency range compared with other experimental techniques used to study the dynamics of supercooled liquids. [37]

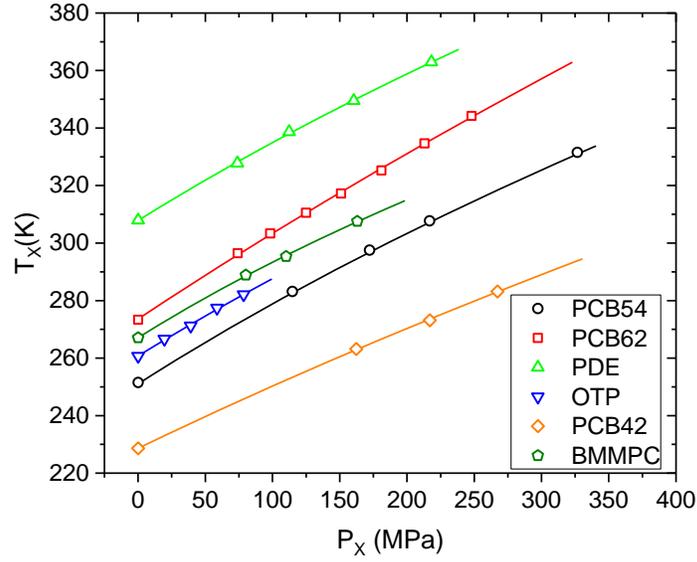

*Figure 1. Temperature $T_X$ versus pressure $P_X$ at constant relaxation time for 6 nonassociated liquids. The points are experimental data and the line are the best fit to the AA equation (eq.(4)). The best-fit parameters are reported in Table 1.*

|       | $T_0$ [K]   | $P_0$ [MPa] | $a$         |
|-------|-------------|-------------|-------------|
| PCB54 | 251.5±0.1   | 350.9±9     | 2.38±0.04   |
| OTP   | 260.7±0.5   | 366±100     | 2.4±1       |
| PCB62 | 273.6±0.5   | 499±87      | 1.77±0.25   |
| PDE   | 307.8±0.6   | 399±95      | 2.6±0.5     |
| PCB42 | 224.55±0.04 | 362±6       | 2.60±0.03   |
| BMMPC | 267±0.5     | 274±76      | 3.3±0.7     |

**Table 1.** Best-fit parameters of the data in Figure 1 to the AA equation (eq.(4)) displayed as solid lines in Figure 1.

For each material, we extracted from the data the pressure dependence of the temperature $T_X$ where $X$ was the dielectric relaxation time τ. Since the change of $T_X$ with pressure increases with increasing τ, for each data set the value of τ chosen was the longest (i.e. closest to the glass transition) for which the largest number of data points was available; for most liquids considered in this study was typically τ=10s. The pressure dependence of $T_X$ for the six liquids is reported in Figure 1 (symbols), together with the best fit (solid lines) to the AA equation (eq. (4)). The best-fit parameters are reported in the Table 1.

In Figure 2 are reported the experimental data (open symbols) of temperature $T_X$ versus the density $\rho_X$ at constant relaxation time. The solid lines in Figure 2 are not a best fit, they are obtained using eq.(6), with the parameters determined from the best-fit of the AA equation (eq. 4) to $T_X(P)$ and the Tait EoS (eq. 5). It is important to notice that both axes in Figure 2 are presented on a logarithmic scale, consequently in this plot the scaling behavior described by eq. (1) with γ=constant would correspond to a linear behavior with a linear coefficient equal to γ. Evidently, the behaviors reported in Figure 2 do not show a strong deviation from a linear behavior and a determination of the pressure dependence of γ is rather difficult, especially without an a priori model describing the nonlinear behavior of $\ln(T_X)$ versus $\ln(\rho_X)$. Instead of determining the pressure dependence of the exponent γ from the deviation from linear behavior of $\ln(T_X)$ versus $\ln(\rho_X)$ data, we use the newly derived equation for γ(P) (eq.(9)) which does not requires any additional data fitting apart from the best-fit parameters obtained from the fit of $\rho(T,P)$ to the Tait EoS and the $T_X(P)$ to the AA equation.

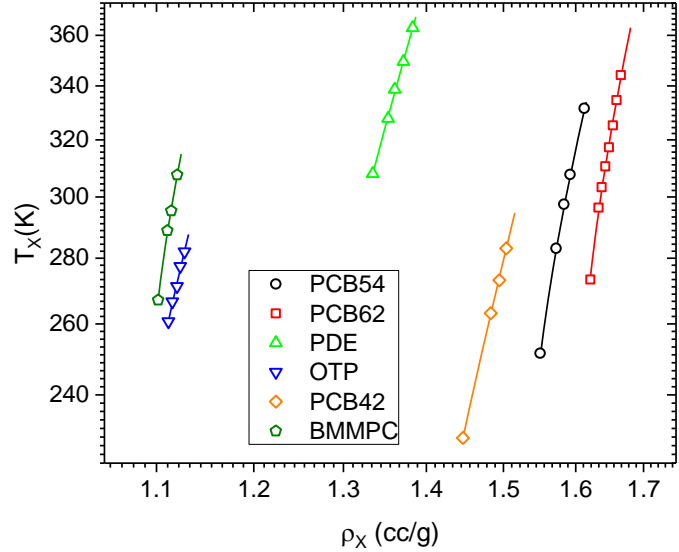

*Figure 2. Log-Log plot of Temperature, $T_X$, versus density, $\rho_X$, at a constant relaxation time for 6 non-associated liquids. The symbols are experimental data and the solid lines are the data calculated using equation (6) using the best fit to the AA equation (eq. (4)) and the Tait EOS (eq.(5)).*

The pressure behavior of the exponent $\gamma(P)$ (obtained using eq.(9)) is reported in Figure 3. The results in Figure 3 clearly shows that for all six liquids the exponent $\gamma$ decreases with pressure. The decrease of $\gamma$ with pressure is more dramatic for liquids having a larger value of $\gamma$ at low pressure, while for materials with $\gamma$ closer to 4 at low pressure the change of $\gamma$ is much smaller. It is important to notice that in all cases, even at the highest pressure, the parameter $\gamma$ remains always larger than 4. This behavior is consistent with that recently observed for DC704, for which $\gamma$ was found to decrease from ~7 to a value close to 4, although over a much larger pressure range (up to P= 0.9 GPa).[12] Comparing the results in Figure 3 with previous values reported of $\gamma$, the methods of obtaining $\gamma$ from a master curve superpositioning dynamic data as a function of $T\rho^{-\gamma}$ or from a linear fit of the data in Figure 2, gives a value close to an average value of $\gamma$.

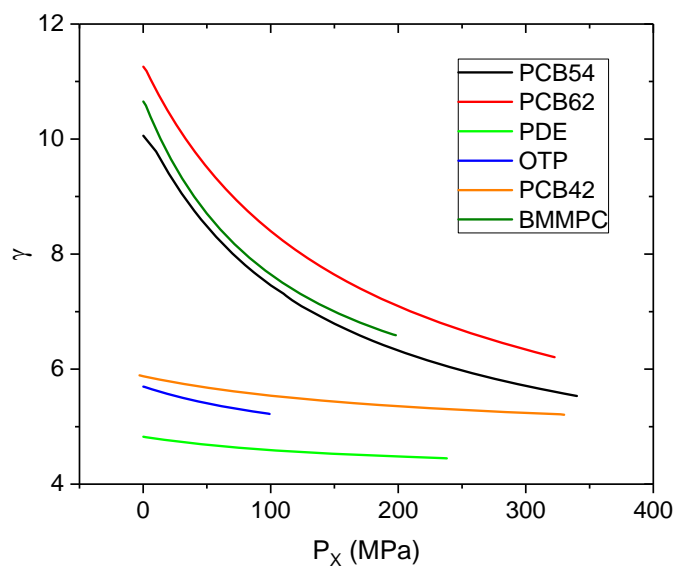

*Figure 3. Pressure dependence of the scaling exponent γ for six nonassociated liquids calculated using equation (9) with the parameters from the EoS and the best-fit of the AA equation to the $T_X$ (P) data (table 1).*

**Discussion**

A limit of the method described above is that the $T_X(P)$ behavior may be better described with different equations than the AA equation. In principle, it could be possible to obtain a different γ(P) behavior. For this reason we analyzed the $T_X(P)$ behavior for the case of PCB62 (since it has the largest number of points) using two nonlinear equations alternative to the AA equation: a quadratic equation $T_X(P) = d_0 + d_1 P + d_2 P^2$ and logarithmic equation $T_X(P) = a_0 \left[1 + a_1 \ln(1 + P/a_2)\right]$ (where $d_n$ and $a_n$ constants).

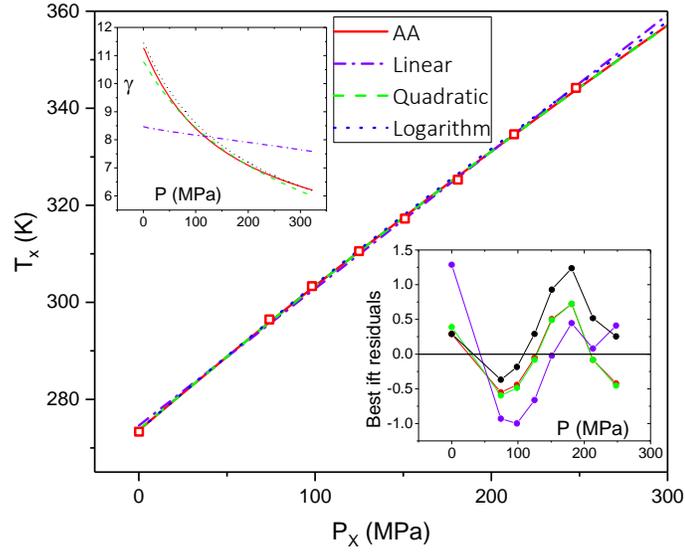

**Figure 4.** *Pressure dependence of $T_X$ for PCB62 (same as in Fig.1) with the best fit to the data using the AA equation and three other equations. A linear equation $T_X(P) = l_0 + l_1 P$ (best-fit parameters $l_0$ =274.6±0.6, $l_1$ =0.282±0.004), a quadratic equation $T_X(P) = d_0 + d_1 P + d_2 P^2$ (best-fit parameters $d_0$ =273.7±0.5, $d_1$ =0.306±0.009, $d_2$ =(-9.5±3.4)×10$^{-5}$) and a logarithmic equation $T_X(P) = a_0 \left[ 1 + a_1 \ln(1 + P/a_2) \right]$ (best-fit parameters $a_0$ = 273.6±0.5, $a_1$ =1316±485, $a_2$ =1.5±0.5).*

The best-fit to the $T_X(P)$ data using these two equations (best-fit parameters are in the Figure 4 caption) are reported in Figure 4 together with the best-fit obtained with the AA equation. From an analysis of the best-fit (lower insert to Figure 4) residuals it is evident that all equations give a good description of the $T_X(P)$ behavior, while larger deviations are observed fitting the data with a linear behavior of $T_X(P)$ with pressure.

The top insert to Figure 4 shows the parameter γ determined by calculating numerically eq.(8) for the four different best-fit equations to $T_X(P)$. We find that as long as the best-fit had a similar deviation from the $T_X(P)$ data (residual are shown in the bottom insert to Figure 4), a similar behavior of γ was found within a deviation of about 0.2. Interestingly, if we analyzed the $T_X(P)$ data using a linear behavior (which has a larger deviation from the data, especially at low pressure, as shown in the bottom insert to Figure 4), the resulting pressure dependence of γ is strongly reduced (top insert to Figure 4). Therefore, a larger deviation from the linear behavior of $T_X(P)$ seems to be an indication of a stronger pressure dependence of γ, although a (reduced) decrease of γ with increasing pressure is observed also in the case of a linear pressure dependence of $T_X(P)$.

The fact, that other equations alternative the AA equation can give a satisfactory description of $T_X(P)$ evidently imposes a limit to the use of eq.(9) to determine the behavior of γ beyond the range of the experimental data. Following the same procedure described above, but substituting the AA equation with a quadratic or logarithmic equation, different equations for the pressure dependence of γ can be obtained and these will certainly have a different behavior at pressure beyond the experimental range of the measurements.

From the results shown above we see that, notwithstanding a constant $\gamma$ was found to give a good superposition of dynamic data and the plot of $\ln(\rho_X)$ versus $\ln(T_X)$ are nearly linear, a significant dependence of $\gamma$ on pressure can be found using eq.(9). In particular, we find that such dependence is larger for materials having $\gamma \gg 4$ and at high pressure the value of $\gamma$ remains larger than 4 like in the case of DC704.[12] This is in contrast with the behavior of associated liquids for which γ was found to increase with pressure approaching the value γ~4 at high pressure.[18] Interestingly, in both cases of associated and nonassociated liquids, even if the pressure behavior is opposite, the condition γ=4 appear to be same high pressure limit.

**Conclusions**

In this study we report a new analytical method to describe the pressure dependence on the exponent γ of the TDS. To use this method it is only necessary to determine the pressure dependence of the temperature at constant $X$, $T_X(P)$ and the EoS.

Using this new method it is possible to determine the pressure dependence of the exponent γ even for liquids for which a constant γ exponent gives a very good superposition of various dynamic properties and an almost linear behavior of $\ln(T_X)$ versus $\ln(\rho_X)$. We find that the previously determined values of γ are close to an average value of the observed γ(P).

For all nonassociated liquids, the exponent γ is found to decrease with pressure. The change of γ(P) is found to be smaller for liquids with γ(P=0) closer to 4, consistent with a high pressure limit of γ~4. This behavior is similar to that found in a recent report on the pressure dependence of the γ exponent for the nonassociated liquid DC704. While it is in contrast with that of associated liquids for which we found γ increases with pressure from γ(0)<4 to γ ~4 at high pressure.

From a theoretical point of view, it has been demonstrated that the TDS behavior (eq.(1)) is predicted in the case for which the intermolecular potential, $U(r)$, is dominated by the repulsive part of the potential and it can be described by an inverse power law behavior $U(r) \propto r^{-n}$. In this case the TDS follows with $\gamma = \frac{n}{3}$ .[38, 39] Molecular dynamic simulations have shown that, in the case of Lennard Jones (LJ) type potentials, in which the attractive part cannot be neglected, the TDS behavior is still verified but with $\gamma > \frac{n}{3}$ (where $n$ is the exponent of the repulsive part), and $3\gamma$ represents an effective slope of the intermolecular potential.[40, 41, 42] Therefore, the variation of the exponent γ with pressure for nonassociated liquids is consistent with a decrease of the effect of the attractive part of the potential with increasing pressure, so that at high pressure the intermolecular potential is dominated by its repulsive part. In particular, since the data are consistent with γ~4 as the high pressure limit of the exponent γ, then the high pressure limit of the potential corresponds to $U(r) \propto r^{-n}$ with n~12. [18] Therefore, current results are consistent with an exponent of the repulsive part of the potential, n~12, for many different nonassociated liquids.

This form for the term of the repulsive potential is evidently consistent with a potential such as the Lennard-Jones 6-12 potential. However, it is not a proof of the validity of the LJ potential, since the high-pressure measurements are relevant for very small intermolecular distances. Our results only indicate that at very small intermolecular distances, the intermolecular potential can be described with a mathematical form that is close to an inverse power law with an exponent close to 12. This is very important since there is currently no other experimental determination of the repulsive part of the potential, and these results may offer an experimental approach to determine a functional form of the repulsive part of the potential that can be used in molecular dynamics simulation.

**Acknowledgments**

This work was supported by the Office of Naval Research. TCR acknowledges an American Society for Engineering Education postdoctoral fellowship.